\documentclass[aps,showpacs,superscriptaddress,preprint]{revtex4}
\usepackage{graphicx}
\usepackage{bm}
\usepackage{amsmath}
\begin{document}
\preprint{USM-TH-115}
\title{Electromagnetic Form Factors of Nucleons in a Light-cone
Diquark Model}
\author{Bo-Qiang Ma}
\email{mabq@phy.pku.edu.cn}
\affiliation{Department of Physics, Peking University, Beijing 100871, China, \\
CCAST (World Laboratory), P.O. Box 8730, Beijing 100080, China,\\
Institute of Theoretical Physics, Academia Sinica, Beijing 100080, China}
\author{Di Qing}
\email{diqing@fis.utfsm.cl}
\author{Iv\'an Schmidt}
\email{ischmidt@fis.utfsm.cl}
\affiliation{Departamento de F\'\i sica, Universidad T\'ecnica Federico
Santa Mar\'\i a, Casilla 110-V, Valpara\'\i so, Chile}

\date{\today}

\begin{abstract}
We investigate the electromagnetic form factors of nucleons within
a simple relativistic quark spectator-diquark model using the
light-cone formalism. Melosh rotations are applied to both quark
and vector diquark. It is shown that the difference between vector
and scalar spectator diquarks reproduces the right electric form
factor of neutrons, and both the form factors
$G_E\left(Q^2\right)$ and $G_M\left(Q^2\right)$ of the proton and
neutron agree with experimental data well up to $Q^2=2
~\rm{GeV}^2$ in this simple model.
\end{abstract}


\maketitle

\section{Introduction}

Since the photon is a particularly clean probe, the
electromagnetic interactions are unique tools for investigating
hadronic physics. Different electromagnetic processes are
described by different transition form factors. For example, in
the case of electron-nucleon elastic scattering processes, the
Sachs form factors $G^N_E$ (electric) and $G^N_M$ (magnetic)
contain all the information. At low momentum transfers, $G^N_E$
and $G^N_M$ are related to the spatial Fourier transform of the
nucleon's charge and magnetization distributions; while at high
momentum transfers, these form factors give important information
about the quark structure within the nucleon and also about the
short-range behavior of the strong interactions. In fact, precise
knowledge of the elastic electromagnetic form factors not only
discloses the internal substructure of nucleons, but is also
important for other processes, such as the experimental
measurement of nucleon strangeness form factors \cite{muller97}
and also for the nucleon spin problem\cite{bro94}.

Theoretically, the very high $Q^2$ behavior of the nucleon form
factors can be well described by perturbative quantum
chromodynamics (QCD) \cite{bro80}. However, as QCD bound states,
and due to the large coupling constant of QCD in the
non-perturbative region, the nucleons have to be explored in
phenomenological QCD models or in lattice QCD. The
non-relativistic constituent quark model is quite successful in
explaining the static properties of baryons, but with increasing
transfer momentum, it must fail in explaining the nucleon form
factors because the intrinsic momentum of the quarks within a
nucleon has the same order of magnitude as the quark mass. Thus,
recently, many studies of nucleon form factors have been extended
to relativistic quark models \cite{dong99}, and also to light-cone
\cite{schlumpf93,cardarelli99} and point-form \cite{wagenbrunn01}
models. These relativistic quark models partially solved the
shortcomings of its non-relativistic counterparts.

The purpose of this work is to show that both the electric and
magnetic form factors of the proton and neutron can be reproduced
in a simple relativistic quark spectator-diquark model which is
formulated in the light-cone frame. This model was originally
proposed in order to study deep inelastic lepton nucleon
scattering \cite{close73}, based on the quark-parton model picture
\cite{feynman69} that deep inelastic scattering is well described
by the impulse approximation, in which the incident lepton
scatters incoherently off a quark in the nucleon, with the
remaining nucleon constituents treated as a quasi-particle
spectator. After taking into account Melosh rotation effects, this
model is in good agreement with the experimental data of polarized
deep inelastic scattering, and the mass difference between the
scalar and vector spectators reproduces the up and down valence
quark asymmetry \cite{ma91}. In this work we extend the
relativistic quark spectator-diquark model in order to investigate
electron-nucleon elastic scattering processes, where the single
quark is the struck constituent and the non-interacting diquark
serves to provide the quantum numbers of the spectator system.
Recently, a covariant quark-diquark model was proposed in order to
describe the nucleon properties \cite{oettel00}, but it was not
formulated in light-cone quantization. The light-cone Fock
representation of composite systems has a number of remarkable
properties. The matrix elements of local operators such as
electromagnetic currents have exact representation in terms of
light-cone wave functions of Fock states \cite{brodsky00}. If one
chooses the special frame \cite{drell70} $q^+=0$ for the
space-like momentum transfer and takes the matrix elements of plus
components of currents, the contribution from pair creation or
annihilation is forbidden and the matrix elements of space-like
currents can be expressed as overlaps of light-cone wave functions
with the same number of Fock constituents.

This paper is organized as follows. In section \ref{sec:model},
the relativistic quark-diquark model formulated in the light-cone
frame is introduced. The specific light-cone form is obtained by
transforming instant into light-cone states using Melosh
rotations. The detailed matrix elements of electromagnetic
currents are given in section \ref{sec:fm}. In section
\ref{sec:result}, we present the results for the nucleon
electromagnetic properties calculated in the light-cone diquark
model. Finally, a brief summary is given in section
\ref{sec:summary}.

\section{The light-cone quark spectator-diquark model}\label{sec:model}

The proton wave function in the conventional $SU\left(6\right)$ quark model
is given by
\begin{equation} \label{eqn:con}
\left.\left.\right|p^\uparrow\right\rangle = \frac{1}{\sqrt{18}}
\left[\left(u^\downarrow d^\uparrow - u^\uparrow d^\downarrow\right)
u^\uparrow +({\rm cyclic\;permutation})\right].
\end{equation}
In the impulse approximation, the incident lepton strikes
incoherently a single constituent quark in the nucleon, with the
remain nucleon constituents treated as a quasi-particle spectator
to provide the other quantum numbers of nucleons. Thus, it is more
convenient to rewrite the wave function of the conventional
$SU\left(6\right)$ quark model in the quark spectator-diquark
form. In fact, in the quark spectator-diquark form, some
non-perturbative effects between the two spectator quarks or other
non-perturbative gluon effects in the nucleon can be effectively
taken into account by the mass of the diquark spectator. Starting
from Eq.~(\ref{eqn:con}), the proton wave function of the
quark-diquark model in the instant form can be written as
\cite{ma91,pavkovi76}
\begin{equation}\label{wav1}
\Psi_p^{\uparrow, \downarrow}\left(qD\right)
=\sin\theta\, \varphi_{V}\,
\left.\left.\right|qV\right\rangle^{\uparrow, \downarrow}
+ \cos\theta\,
\varphi_{S}\, \left.\left.\right|qS\right\rangle^{\uparrow, \downarrow},
\end{equation}
with
\begin{eqnarray}\label{wav2}
\left.\left.\right|qV\right\rangle^{\uparrow, \downarrow}
&=&\pm \frac{1}{3}\left[V^0\left(ud\right)u^{\uparrow, \downarrow}-
\sqrt{2}V^{\pm 1}\left(ud\right)u^{\downarrow, \uparrow}
-\sqrt{2}V^0\left(uu\right)d^{\uparrow, \downarrow}
+2V^{\pm 1}\left(uu\right)d^{\downarrow, \uparrow}\right], \nonumber\\
\left.\left.\right|qS\right\rangle^{\uparrow, \downarrow}
&=&S\left(ud\right)u^{\uparrow, \downarrow},
\end{eqnarray}
where $\theta$ is the mixing angle that breaks SU(6) symmetry at
$\theta\neq \pi /4$, $V^{s_z}(q_{1}q_{2})$ is the $q_1 q_2$
instant form vector diquark Fock state with third spin component
$s_z$, $S(ud)$ is the $ud$ scalar diquark Fock state, and
$\varphi_{D}$ stands for the momentum space wave function of the
quark-diquark, with $D$ representing the vector ($V$) or scalar
($S$) diquarks. In this paper we choose the $SU\left(6\right)$
symmetry case $\theta=\pi /4$. The neutron wave function can be
obtained simply by exchanging the $u$ and $d$ quarks.

Since the light-cone Fock representation of composite systems has
a number of remarkable properties, we will calculate the
electromagnetic properties of nucleons in this formalism. The
light-cone momentum space wave function of the quark-spectator is
assumed to be a harmonic oscillator wave function (the
Brodsky-Huang-Lepage (BHL) prescription \cite{brodsky81,Huang94})
\begin{equation}
\varphi_D\left(x,\bm{k}_{\perp}\right) = A_D exp\left\{ -\frac{1}{8\beta_D^2}
\left[\frac{m_q^2+\bm{k}_{\perp}^2}{x}+\frac{m_D^2+
\bm{k}_{\perp}^2}{1-x}\right]\right\},
\end{equation}
where $m_q$ and $m_D$ are the masses for the quark $q$ and
spectator $D$ respectively, $x$ and ${\bm k}_\perp$ are the
light-cone momentum fraction and internal transversal momentum of
quarks respectively, and $\beta_D$ is the harmonic oscillator
scale parameter.

The spin part of the light-cone wave functions of nucleons can be
obtained by transforming instant into light-cone states using
Melosh rotations. For a spin-$\frac{1}{2}$ particle, the Melosh
transformations are known to be \cite{melosh74}
\begin{eqnarray}\label{eqn:melosh}
\chi^{\uparrow}_T &=& w \left[\left(k^+ + m\right)
\chi^{\uparrow}_F - k^R \chi^{\downarrow}_F\right],
\nonumber \\
\chi^{\downarrow}_T &=& w \left[\left(k^+ + m\right)
\chi^{\downarrow}_F - k^L \chi^{\uparrow}_F\right],
\end{eqnarray}
where $\chi_T$ and $\chi_F$ are instant and light-cone spinors
respectively,
$w=\left[2k^+\left(k^0+m_q\right)\right]^{-\frac{1}{2}}$,
$k^{R,L}=k^1\pm i k^2$, and $k^+ = k^0+ k^3$. In this work, for
simplicity we treat the diquark as a point particle. The scalar
diquark does not transform, since it has zero spin. For the
spin-$1$ vector diquark, the Melosh transformations are given by
Ahluwalia and Sawicki as \cite{ahluwalia93}
\begin{eqnarray}\label{eqn:melosh1}
V^1_T &=& w^2 \left[\left(k^+ +m\right)^2 V^1_F
-\sqrt{2}\left(k^+ +m\right)k^R V^0_F + {k^R}^2 V^{-1}_F\right],\nonumber\\
V^0_T &=& w^2 \left[\sqrt{2}\left(k^+ +m\right)k^L V^1_F
+ 2\left(\left(k^0+m\right)k^+-k^R k^L\right)V^0_F-\sqrt{2}\left(k^+
+m\right)k^R V^{-1}_F\right], \nonumber\\
V^{-1}_T &=& w^2 \left[{k^L}^2 V^1_F+\sqrt{2}\left(k^+ +m\right)k^L
V^0_F + \left(k^+ +m\right)^2V^{-1}_F\right].
\end{eqnarray}
Here, $V_T$ and $V_F$ are the instant and light-cone spin-$1$
spinors respectively, which are constructed within the
Weinberg-Soper formalism \cite{weinberg64}. Ahluwalia and Sawicki
\cite{ahluwalia93} used this formalism in order to construct
explicit hadronic spinors for arbitrary spin in the light-cone
form, and obtained general Melosh rotations for spinors of
arbitrary spin. In the case of spin-$\frac{1}{2}$, these spinors
are in agreement with the results of Lepage and Brodsky
\cite{bro80}, and also the Melosh rotations are the same as in
Eq.~(\ref{eqn:melosh}), which were introduced originally by
Melosh. The Melosh rotations for spin-$1$ particles have a similar
form as Eq.~(\ref{eqn:melosh1}), if the spin-$1$ particle is
treated as a composite state of two spin-$\frac{1}{2}$ particles.

\section{The calculation of electromagnetic form factors}\label{sec:fm}

The nucleon Sachs form factors are defined as combinations of the
Dirac and Pauli form factors \cite{sachs62}
\begin{eqnarray}
G_E^N\left(Q^2\right) &=& F_1^N\left(Q^2\right)-\frac{Q^2}{4M^2}
F_2^N\left(Q^2\right), \nonumber \\
G_M^N\left(Q^2\right) &=& F_1^N\left(Q^2\right)+F_2^N\left(Q^2\right).
\end{eqnarray}
As a spin-$\frac{1}{2}$ composite system, the nucleon has Dirac
and Pauli form factors $F_1\left(Q^2\right)$ and
$F_2\left(Q^2\right)$ defined by
\begin{equation}
\left\langle P^\prime\left|J^\mu\left(0\right)\right|P\right\rangle
=\bar{u}\left(P^\prime\right)\left[F_1^N\left(Q^2\right)\gamma^\mu
+F_2^N\left(Q^2\right)\frac{i\sigma^{\mu\nu}q_\nu}{2M}\right]u\left(P\right),
\end{equation}
where $q^\mu=\left(P^\prime-P\right)^\mu$ is four-momentum
transfer, $Q^2=-q^2$, and $u\left(P\right)$ is the nucleon spinor.
In the light-cone formalism, it is well known that the Dirac and
Pauli form factors are equal to the helicity-conserving and
helicity-flip matrix elements of the plus component of
electromagnetic current operators \cite{brodsky80}:
\begin{eqnarray}
\left\langle P^\prime,\uparrow\left|\frac{J^+\left(0\right)}{2P^+}
\right| P,\uparrow\right\rangle &=&
F_1^N\left(Q^2\right),\nonumber\\ \left\langle
P^\prime,\uparrow\left|\frac{J^+\left(0\right)}{2P^+} \right|
P,\downarrow\right\rangle &=& -(q_1- i q_2)
\frac{F_2^N\left(Q^2\right)}{2M}.
\end{eqnarray}

In the $Q^2\rightarrow 0$ limit, the nucleon magnetic form factor
is just the magnetic moment of the nucleon:
\begin{equation}
\mu_N= G_M^N\left(0\right).
\end{equation}
Other important static electromagnetic properties of nucleons are
the charge and magnetic radii, which can be obtained via the
$Q^2\rightarrow 0$ limit of the electromagnetic form factors as:
\begin{eqnarray}
r_{EN}^2 &=& -6\, \lim_{Q^2\rightarrow 0}\frac{dG_E^N\left(Q^2\right)}
{dQ^2}, \nonumber \\
r_{MN}^2 &=& -6\, \lim_{Q^2\rightarrow 0}\frac{1}{\mu_N}
\frac{dG_M^N\left(Q^2\right)} {dQ^2}.
\end{eqnarray}

In these calculations, we choose the Drell-Yan assignment
\cite{drell70}:
\begin{eqnarray}
q&=&\left(q^+,q^-,\bm{q}_{\perp}\right) = \left(0,\frac{-q^2}{P^+},
\bm{q}_{\perp}\right), \nonumber \\
P&=&\left(P^+,P^-,\bm{P}_{\perp}\right) = \left(P^+,\frac{M^2}{P^+},
\bm{0}_{\perp}\right).
\end{eqnarray}
As pointed out in the introduction, the matrix elements of
space-like currents can be expressed as overlaps of light-cone
wave functions with the same number of Fock constituents. In
particular, for the form factors we have
\begin{eqnarray}\label{eqn:overlap}
F_1^N\left(Q^2\right)&=&\sum_a\int\frac{d^2\bm{k}_\perp dx}{16\pi^3}
\sum_j e_j\psi^{\uparrow\star}_a\left(x_i,\bm{k}^\prime_{\perp i},
\lambda_i\right) \psi^{\uparrow}_a\left(x_i,\bm{k}_{\perp i},
\lambda_i\right), \nonumber \\
-q^L\frac{F_2^N\left(Q^2\right)}{2M}
&=&\sum_a\int\frac{d^2\bm{k}_\perp dx}{16\pi^3}
\sum_j e_j\psi^{\uparrow\star}_a\left(x_i,\bm{k}^\prime_{\perp i},
\lambda_i\right) \psi^{\downarrow}_a\left(x_i,\bm{k}_{\perp i},
\lambda_i\right),
\end{eqnarray}
where $e_j$ is the charge of struck constituents, $\psi_a
\left(x_i,\bm{k}_{\perp i},\lambda_i\right)$ is the light-cone
Fock expansion wave function, and $\lambda_i$, $x_i$ and
$\bm{k}_{\perp i}$ are the spin projections along the quantization
direction, light-cone momentum fractions and relative momentum
coordinates of the QCD constituents, respectively. Here, for the
final state light-cone wave function, the relative momentum
coordinates are
\begin{equation}
\bm{k}^\prime_{\perp i}=\bm{k}_{\perp i}+\left(1-x_i\right)\bm{q}_{\perp}
\end{equation}
for the struck quark and
\begin{equation}
\bm{k}^\prime_{\perp i}=\bm{k}_{\perp i}-x_i\bm{q}_{\perp}
\end{equation}
for each spectator. This treatment is similar to that for the pion
form factor with a quark-antiquark pair as constituents in the
light-cone formalism \cite{Ma93}.

Thus, in terms of the wave functions Eqs.~(\ref{wav1}, \ref{wav2},
\ref{eqn:melosh}) and (\ref{eqn:melosh1}), it is easy to obtain
the expressions for the Dirac and Pauli form factors in our
light-cone quark model. Explicit expressions for the proton form
factors are:
\begin{eqnarray}
F_1^p\left(Q^2\right)&=&3\int\frac{d^2 k_\perp dx}{16\pi^3}\frac{2}{3}
\cos^2\theta w^\prime_q w_q \left[\left(k^{\prime +}_q+m_q\right)\left(
k^+_q+m_q\right)+k^{\prime L}_\perp k^R_\perp\right]\varphi_S\left(x,
\bm{k}^\prime_\perp\right)\varphi_S\left(x,\bm{k}_\perp\right), \nonumber \\
F_2^p\left(Q^2\right)&=&\frac{6M}{-q^L}\int
\frac{d^2 k_\perp dx}{16\pi^3}\frac{2}{3}\cos^2\theta w^\prime_q w_q
\left[\left(k^{\prime +}_q+m_q\right)k^{L}_\perp - \left(
k^+_q+m_q\right)k^{\prime L}_\perp\right]\nonumber \\
&& \,\, \times\varphi_S\left(x,
\bm{k}^\prime_\perp\right)\varphi_S\left(x,\bm{k}_\perp\right).
\end{eqnarray}
It is interesting to notice that in the proton case the
contributions of the vector diquark cancel each other exactly due
to the spin-flavor structure of the proton wave functions. In the
case of neutrons, the expressions are similar, but the
contributions of vector diquarks do not cancel. For completion, we
also present them in the Appendix.

\section{Results and discussions}\label{sec:result}

In our model there are five parameters: the quark mass $m_q$, the
scalar and vector diquark masses $m_S$ and $m_V$, and the harmonic
oscillator scale parameter $\beta_S$ and $\beta_V$, which are
fixed by six static electromagnetic properties: the charge radii,
magnetic radii and magnetic moments of both proton and neutron.
This is given in Table \ref{tab:parameters}. As we already
noticed, in the proton case the vector diquark contributions to
the form factors cancel each other exactly. Therefore, the three
static electromagnetic properties of protons can fix completely
the quark mass and the parameters corresponding to the scalar
diquark. The other parameters are fixed by the neutron static
properties. We calculate the static electromagnetic properties and
form factors of nucleons using three different sets of parameters
in order to show their dependence on the difference between the
scalar and vector diquarks. We find that the electromagnetic
properties of neutrons, specially the charge radius, are sensitive
to the values of the parameters. When the value of the vector
diquark harmonic oscillator scale parameter is less than that of
the scalar diquark, we obtain a much larger positive squared
charge radius for the neutron, although the neutron magnetic
moment is in good agreement with experiment. In our calculations,
the quark mass $m_q=0.220\,{\rm GeV}$ is much less than that used
in the non-relativistic quark model $(m_q\sim M_N/3=0.313\, {\rm
GeV})$. This is due to the fact that although in the
non-relativistic quark model the proton magnetic moment has a
simple relation with the quark mass, $\mu_p=\frac{M_N}{m_q}\sim
3$, in relativistic quark models, for example our model, this
relation is more complicated, and it depends on both the mass and
the intrinsic momentum of quarks within the nucleon. In fact, the
value of the quark mass used in our model is comparable with that
used in other relativistic quark model \cite{capstick86}. The
model parameters used here are slightly different from those used
in previous calculations of valence quark distributions, but these
previous results are not sensitive to the actual values of the
parameters and therefore the present parameters will give similar
results.

The static electromagnetic properties of nucleons are shown in
Table \ref{tab:properties} for three different sets of parameters.
In the first set we use the same mass and scale parameters for the
scalar and vector diquarks in order to keep the $SU\left(6\right)$
symmetry of nucleon wave functions, but get nearly zero charge
radius for the neutron and also the proton anomalous magnetic is
close to that of the neutron. This choice of parameters is
compatible with the naive $SU\left(6\right)$ symmetric quark
model. However, the experimental negative squared charge radius of
neutrons as well as the ratio $\mu_p/\mu_n=-1.46$ indicate that
the $SU\left(6\right)$ symmetry is broken. In our model the
$SU\left(6\right)$ breaking effect is introduced by the difference
between the scalar and vector diquarks parameters, as it is shown
with the third set of parameters. Except for the fact that the
neutron magnetic moment is slightly off, the results of set III
are in good agreement with experiment, through the difference
between the scalar and vector diquarks parameters, which breaks
the $SU\left(6\right)$ symmetry and gives the right neutron charge
radius. Although the Melosh rotations already break the
$SU\left(6\right)$ symmetry of nucleon wave functions, it is not
enough to produce the right neutron charge radius
\cite{cardarelli99}. In set II, we try to fix the neutron magnetic
moment. However, we obtain a bad result for the neutron charge
radius.

\begin{figure}
\includegraphics{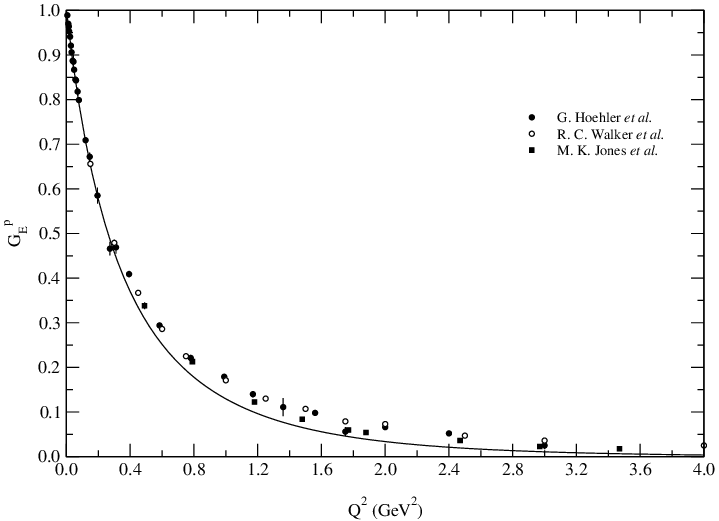}
\caption{\label{fig:pge}Electric form factor of protons. The experimental
data are from \cite{hoehler76,walker94,jones00}}
\end{figure}

\begin{figure}
\includegraphics{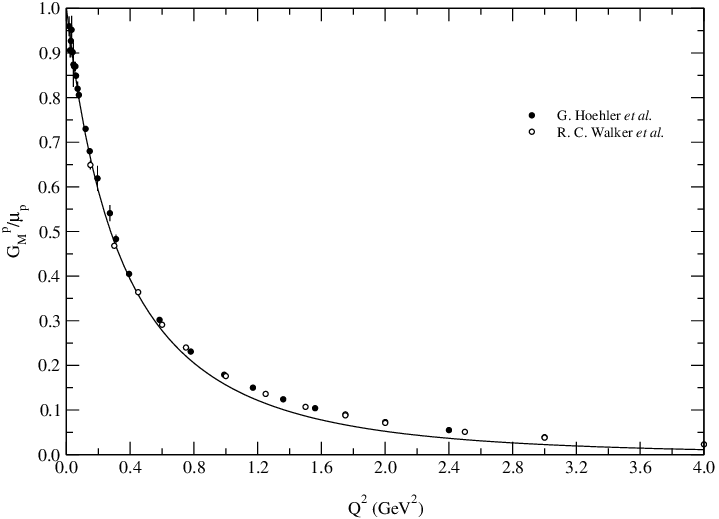}
\caption{\label{fig:pgm}Magnetic form factor of protons. The experimental
data are from \cite{hoehler76,walker94}.}
\end{figure}

\begin{figure}
\includegraphics{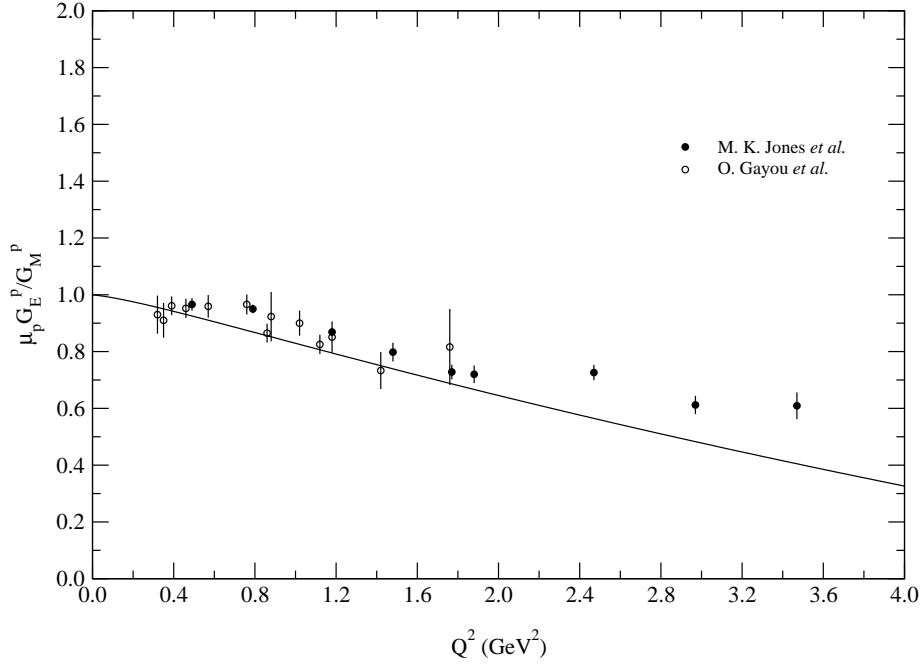}
\caption{\label{fig:pgem}The ratio $\left(\mu_pG_E/G_M\right)$ of protons
compared with the data from ref. \cite{jones00}.}
\end{figure}

\begin{figure}
\includegraphics{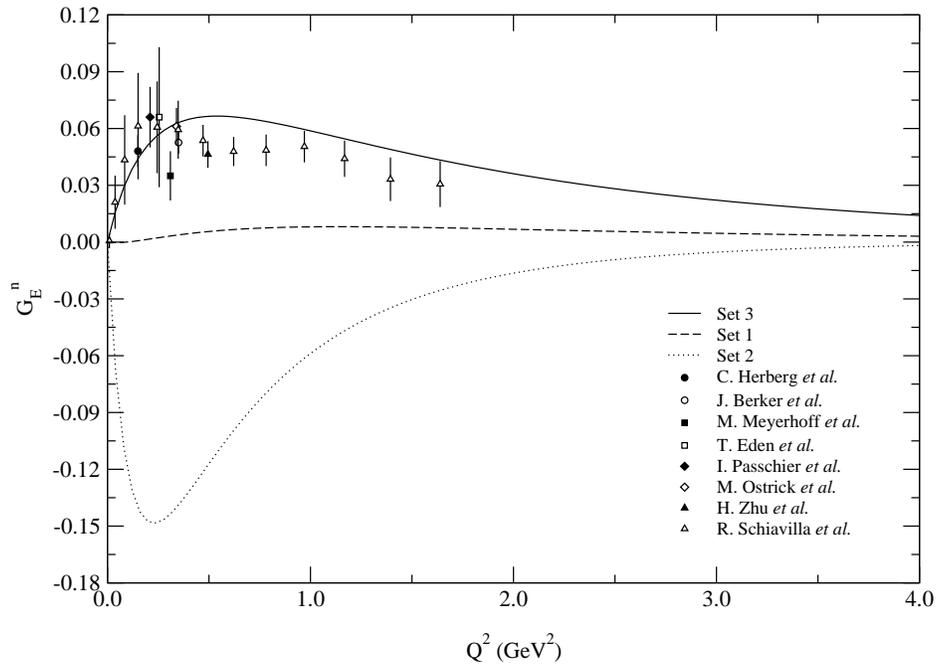}
\caption{\label{fig:nge}Electric form factor of neutrons. The experimental
data are from \cite{meyerhoff94}.}
\end{figure}

\begin{figure}
\includegraphics{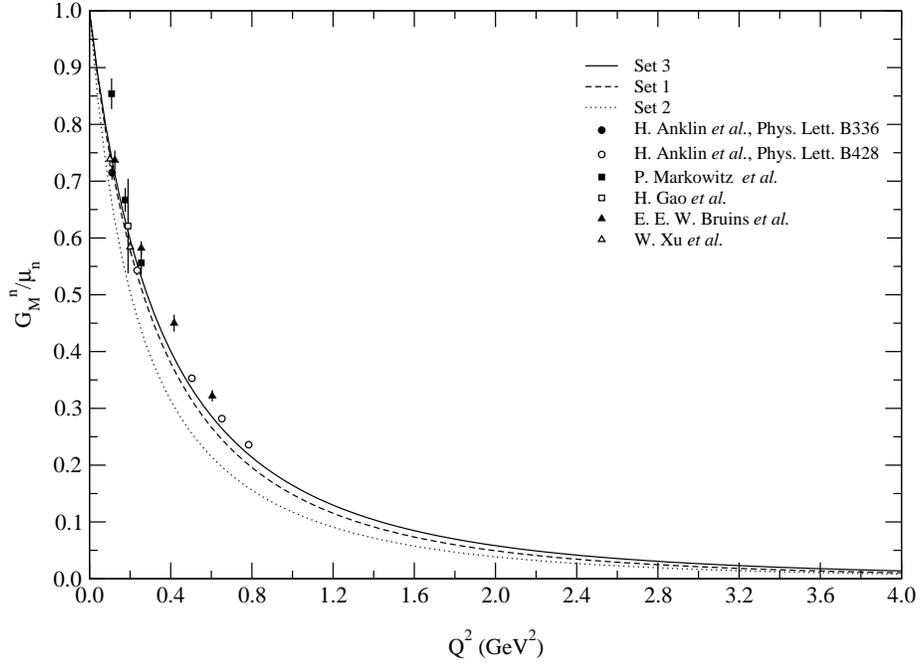}
\caption{\label{fig:ngm}Magnetic form factor of neutrons. The experimental
data are from \cite{anklin94}.}
\end{figure}

In Figs.~\ref{fig:pge} and \ref{fig:pgm} we show the
electromagnetic form factors of protons up to $Q^2=4\, {\rm
GeV}^2$. Since as pointed out in Ref. \cite{jones00} the results
for the electric form factor of protons from Ref.
\cite{hoehler76,walker94} are too high, in Fig ~\ref{fig:pge} we
also present the results from Ref. \cite{jones00} for $G_E^p$,
which are obtained by assuming that the proton magnetic form
factor $G_M^p/\mu_p$ satisfies a dipole behaviour
$G_D=1/\left[1+Q^2/0.71\right]^{2}$. Both form factors are in good
agreement with experiment even up to $Q^2=4\, {\rm GeV}^2$.
However, because of the harmonic oscillator wave functions used in
our calculations, it is important to keep in mind that we expect
our model to be valid essentially for $Q^2$ up to $2\,{\rm
GeV}^2$. Recently, the JLab data \cite{jones00} showed that the
$Q^2$ dependence of $G_E^p$ and of $G_M^p$ is significantly
different, in contrast to the non-relativistic quark model
prediction, in which both have the same behavior. In the
light-cone frame, however, after taking into account the Melosh
rotations, the $Q^2$ dependence is naturally different because the
Dirac form factor is obtained from the helicity-conserving matrix
element of the plus component of electromagnetic current operator
while the Pauli form factor is obtained from the helicity-flip
matrix element. Our results for the ratio between $G_E^p$ and
$G_M^p$, together with the JLab data, are shown in
Fig.~\ref{fig:pgem}, and the neutron electromagnetic form factors
are shown in Figs.~\ref{fig:nge} and \ref{fig:ngm}. In contrast to
the proton case, the experimental data for the neutron has fewer
points, and our results also agree with the available experimental
data. Once again it is the difference between the scalar and
vector diquarks which breaks the $SU\left(6\right)$ symmetry and
reproduces the non-zero neutron electromagnetic form factor.

\section{summary}\label{sec:summary}

We have calculated the electromagnetic form factors of nucleons in
a simple quark spectator-diquark model which is formulated in the
light-cone formalism. The wave functions of the model are obtained
by transforming the instant states into the light-cone states
using Melosh rotations which are applied to both the quark and
vector diquark. The model parameters are fixed by the six static
electromagnetic properties of nucleons. We find that the
calculated form factors are in good agreement with the
experimental data, and that the ratio between the electric and
magnetic form factors of protons also agrees with recent JLab
data. The difference between the scalar and vector diquarks breaks
the $SU\left(6\right)$ symmetry of the nucleon wave functions and
it is essential in order to reproduce the correct electromagnetic
nucleon properties.
This framework is certainly applicable to calculate other
processes such as nucleon-delta transitions and nucleon axial form
factors, and would therefore be interesting to extend and check it
in its application to these processes and also to reactions
involving other baryons.

\acknowledgments{This work is partially supported by National
Natural Science Foundation of China under Grant Numbers 19975052£¬
10025523, and 90103007, by Fondecyt (Chile) under project 3000055 and Grant
Number 8000017.}

\appendix*
\section{}
Using Eqs. (\ref{wav1},\ref{wav2},\ref{eqn:melosh}, \ref{eqn:melosh1})
and (\ref{eqn:overlap}), we obtain the explicit expressions of neutron
form factors as:
\begin{eqnarray}
F_1^n\left(Q^2\right)&=&3\int\frac{d^2 k_\perp dx}{16\pi^3}w^\prime_q w_q
\left\{\frac{1}{9} \sin^2\theta \left\{\left[\left(k^{\prime +}_q+
m_q\right)\left(k^+_q+m_q\right)+k^{\prime L}_\perp k^R_\perp\right]
O_{V^{0,0}}\right.\right.\nonumber \\
&&-\left.\left.\sqrt{2}\left[\left(k^{\prime +}_q+ m_q\right)k^L_\perp
-\left(k^+_q+m_q\right)k^{\prime L}_\perp\right]O_{V^{0,1}}
\right.\right.\nonumber \\
&&-\left.\left.\sqrt{2}\left[\left(k^+_q+m_q\right)k^{\prime R}_\perp
-\left(k^{\prime +}_q+ m_q\right)k^R_\perp\right]O_{V^{1,0}}
\right.\right.\nonumber \\
&& +\left.\left.2\left[\left(k^{\prime +}_q+
m_q\right)\left(k^+_q+m_q\right)+k^{\prime R}_\perp k^L_\perp\right]
O_{V^{1,1}}\right\}\varphi_V\left(x,\bm{k}^\prime_\perp\right)
\varphi_V\left(x,\bm{k}_\perp\right)\right.\nonumber \\
&&\left.-\frac{1}{3}
\cos^2\theta \left[\left(k^{\prime +}_q+m_q\right)\left(
k^+_q+m_q\right)+k^{\prime L}_\perp k^R_\perp\right]\varphi_S\left(x,
\bm{k}^\prime_\perp\right)\varphi_S\left(x,\bm{k}_\perp\right)\right\},
\end{eqnarray}
and
\begin{eqnarray}
F_2^n\left(Q^2\right)&=&\frac{6M}{-q^L}\int \frac{d^2 k_\perp dx}{16\pi^3}
w^\prime_q w_q\left\{-\frac{1}{9}\sin^2\theta\left\{\left[\left(k^{\prime +}_q
+m_q\right)k^L_\perp - \left(k^+_q+m_q\right)k^{\prime L}_\perp\right]
O_{V^{0,0}}\right.\right.\nonumber \\
&&\left.\left. -\sqrt{2}\left[\left(k^{\prime +}_q+
m_q\right)\left(k^+_q+m_q\right)+k^{\prime L}_\perp k^R_\perp\right]
O_{V^{0,-1}}\right.\right.\nonumber \\
&&\left.\left. -\sqrt{2}\left[\left(k^{\prime +}_q+
m_q\right)\left(k^+_q+m_q\right)+k^{\prime R}_\perp k^L_\perp\right]
O_{V^{1,0}}\right.\right.\nonumber \\
&&\left.\left. +2\left[\left(k^+_q+m_q\right)k^{\prime R}_\perp
-\left(k^{\prime +}_q+ m_q\right)k^R_\perp\right]O_{V^{1,-1}}
\right\}\varphi_V\left(x,\bm{k}^\prime_\perp\right)
\varphi_V\left(x,\bm{k}_\perp\right) \right.\nonumber \\
&&\left. -\frac{1}{3}\cos^2\theta
\left[\left(k^{\prime +}_q+m_q\right)k^{L}_\perp - \left(
k^+_q+m_q\right)k^{\prime L}_\perp\right]\varphi_S\left(x,
\bm{k}^\prime_\perp\right)\varphi_S\left(x,\bm{k}_\perp\right)
\right\},
\end{eqnarray}
where
\begin{eqnarray}
O_{V^{0,0}}&=&{w^\prime}^2_D w^2_D\left[4\left(k^{\prime +}_D+m_D\right)
\left(k^+_D+m_D\right){\bm k}^\prime_\perp\cdot{\bm k}_\perp
+\left(\left(k^{\prime +}_D+m_D\right)^2-{{\bm k}^\prime}^2_\perp
\right)\right. \nonumber\\
&& \left. \times\left(\left(k^+_D+m_D\right)^2-{\bm k}^2_\perp\right)\right],
\nonumber\\
O_{V^{0,1}}&=&{w^\prime}^2_D w^2_D\left[-\sqrt{2}\left(k^{\prime +}_D+m_D\right)
\left(k^+_D+m_D\right)^2 k^{\prime R}_\perp + \sqrt{2}\left(\left(
k^{\prime +}_D+m_D\right)^2-{{\bm k}^\prime}^2_\perp\right)
\right.\nonumber \\
&& \left.\times\left(k^+_D+m_D\right)k^R_\perp +\sqrt{2}
\left(k^{\prime +}_D+m_D\right)k^{\prime L}_\perp {k^R_\perp}^2
\right],\nonumber\\
O_{V^{1,0}}&=&{w^\prime}^2_D w^2_D\left[-\sqrt{2}\left(
k^{\prime +}_D+m_D\right)^2
\left(k^+_D+m_D\right)k^L_\perp + \sqrt{2}\left(k^{\prime +}_D+m_D\right)
\right.\nonumber \\
&& \left.\times\left(\left(k^+_D+m_D\right)^2-{\bm k}^2_\perp\right)
k^{\prime L}_\perp + \sqrt{2}\left(k^+_D+m_D\right){k^{\prime L}_\perp}^2
k^R_\perp \right],\nonumber\\
O_{V^{1,1}}&=&{w^\prime}^2_D w^2_D\left[\left(k^{\prime +}_D+m_D\right)^2
\left(k^+_D+m_D\right)^2 + 2\left(k^{\prime +}_D+m_D\right)
\left(k^+_D+m_D\right) \right. \nonumber \\
&& \left.\times k^{\prime L}_\perp k^R_\perp +{k^{\prime L}_\perp}^2
{k^R_\perp}^2\right], \nonumber\\
O_{V^{0,-1}}&=&{w^\prime}^2_D w^2_D\left[-\sqrt{2}\left(
k^{\prime +}_D+m_D\right)
k^{\prime R}_\perp {k^L_\perp}^2 - \sqrt{2}\left(\left(
k^{\prime +}_D+m_D\right)^2-{{\bm k}^\prime}^2_\perp\right)
\left(k^+_D+m_D\right)k^L_\perp \right. \nonumber\\
&& \left. +\sqrt{2}\left(k^{\prime +}_D+m_D\right)\left(k^+_D+m_D\right)^2
k^{\prime L}_\perp\right],
\end{eqnarray}
and
\begin{eqnarray}
O_{V^{0,-1}}&=&{w^\prime}^2_D w^2_D\left[\left(k^{\prime +}_D+m_D\right)^2
{k^L_\perp}^2 - 2\left(k^{\prime +}_D+m_D\right)\left(k^+_D+m_D\right)
\right. \nonumber\\
&& \left. \times k^{\prime L}_\perp k^L_\perp +\left(k^+_D+m_D\right)^2
{k^{\prime L}_\perp}^2\right]
\end{eqnarray}
come from the Melosh rotations of vector diquarks, in which the
superscript $\left(i,j\right)$ of $V$ indicates the third spin
component of initial and final state vector diquarks.


\newpage
\begin{table}
\caption{\label{tab:parameters}The three sets of parameters used in
the calculations.}
\begin{ruledtabular}
\begin{tabular}{rrrr}
 & Set I & Set II & Set III  \\
\hline
$m_q$ (GeV) & 0.220 & 0.220 & 0.220 \\
$m_{s}$ (GeV) & 0.500 & 0.500 & 0.500 \\
$m_{v}$ (GeV) & 0.500 & 0.700 & 0.700 \\
$\beta_{s}$ (GeV) & 0.250 & 0.250 & 0.250 \\
$\beta_{v}$ (GeV) & 0.250 & 0.170 & 0.330 \\
\end{tabular}
\end{ruledtabular}
\end{table}

\begin{table}
\caption{\label{tab:properties}The static electromagnetic properties
of nucleons for the three sets of parameters.}
\begin{ruledtabular}
\begin{tabular}{lrrrr}
        & Set I & Set II & Set III & Experimental \\
\hline
$\left(r_p\right)_{el} \left({\rm fm}\right)$ & 0.864 & 0.864 & 0.864 &
0.883(14)\cite{melnikov00},0.880(15)\cite{rosenfelder00} \\
$\left(r^2_n\right)_{el} \left({\rm fm}^2\right)$ & 0.003 & 0.514 & $-$0.110 &
$-$0.113(7)\cite{kopecky95} \\
$\left(r_p\right)_{mag} \left({\rm fm}\right)$ & 0.853 & 0.853 & 0.853 &
0.843(13)\cite{hoehler76} \\
$\left(r_n\right)_{mag} \left({\rm fm}\right)$ & 0.873 & 1.016 & 0.847 &
0.840(42)\cite{hoehler76} \\
$\mu_p$ & 2.797 & 2.797 & 2.797 & 2.793\cite{pdg00} \\
$\mu_n$ & $-$1.840 & -1.903 & -1.664 & $-$1.913\cite{pdg00} \\
\end{tabular}
\end{ruledtabular}
\end{table}

\end{document}